
%
%
\newcount\equationno      \equationno=0
\newcount\itemno  \itemno=0
\newtoks\chapterno \xdef\chapterno{}
\newdimen\tabledimen  \tabledimen=\hsize
%
%
%
\def\eqn{\eqno\eqname}
\def\eqname#1{\global \advance \equationno by 1 \relax
\xdef#1{{\noexpand{\rm}(\chapterno\number\equationno)}}#1}
%
%
%
%
\def\table#1#2{\tabledimen=\hsize \advance\tabledimen by -#1\relax
\divide\tabledimen by 2\relax\vskip 1pt
\moveright\tabledimen\vbox{\tabskip=1em plus 4em minus 0.9em
\halign to #1{#2}} }
%
%
%
\def\tablec#1#2{\tabledimen=\hsize \advance\tabledimen by -#1\relax
\divide\tabledimen by 2\relax\vskip 1pt
\moveright\tabledimen\vbox{\tabskip=1em plus 4em minus 0.9em
\halign to #1{&\hfil##\hfil\crcr #2}} }

%
\def\tablev#1#2#3{\tabledimen=\hsize \advance\tabledimen by -#1\relax
\divide\tabledimen by 2\relax\vskip 1pt
\moveright\tabledimen\vbox{\offinterlineskip\tabskip=0em
\halign to #1{\vrule##\tabskip=1em plus 4em minus 0.9em&\strut#2
&\vrule##\tabskip 0 em\crcr #3}} }
%

%
\def\boxit#1{\vbox{\hrule\hbox{\vrule\kern3pt
\vbox{\kern3pt#1\kern3pt}\kern3pt\vrule}\hrule}}
%
\def\sqr#1#2{{\vcenter{\vbox{\hrule height .#2pt
 \hbox{\vrule width .#2 pt height #1 pt \kern#1pt\vrule width .#2pt}
 \hrule height .#2pt}}}}

%
\def\beginitem{\global \advance \itemno by 1 \relax
\item{\bf \number\itemno.\hfil}}
%


%
\def\utw{\smash{\rlap{\lower5pt\hbox{$\sim$}}}}
\def\udtw{\smash{\rlap{\lower6pt\hbox{$\approx$}}}}

%

%



%



\def\bbbc{{\mathchoice {\setbox0=\hbox{$\displaystyle\rm C$}\hbox{\hbox
to0pt{\kern0.4\wd0\vrule height0.9\ht0\hss}\box0}}
{\setbox0=\hbox{$\textstyle\rm C$}\hbox{\hbox
to0pt{\kern0.4\wd0\vrule height0.9\ht0\hss}\box0}}
{\setbox0=\hbox{$\scriptstyle\rm C$}\hbox{\hbox
to0pt{\kern0.4\wd0\vrule height0.9\ht0\hss}\box0}}
{\setbox0=\hbox{$\scriptscriptstyle\rm C$}\hbox{\hbox
to0pt{\kern0.4\wd0\vrule height0.9\ht0\hss}\box0}}}}
\def\bbbe{{\mathchoice {\setbox0=\hbox{\smalletextfont e}\hbox{\raise
0.1\ht0\hbox to0pt{\kern0.4\wd0\vrule width0.3pt height0.7\ht0\hss}\box0}}
{\setbox0=\hbox{\smalletextfont e}\hbox{\raise
0.1\ht0\hbox to0pt{\kern0.4\wd0\vrule width0.3pt height0.7\ht0\hss}\box0}}
{\setbox0=\hbox{\smallescriptfont e}\hbox{\raise
0.1\ht0\hbox to0pt{\kern0.5\wd0\vrule width0.2pt height0.7\ht0\hss}\box0}}
{\setbox0=\hbox{\smallescriptscriptfont e}\hbox{\raise
0.1\ht0\hbox to0pt{\kern0.4\wd0\vrule width0.2pt height0.7\ht0\hss}\box0}}}}
\def\bbbq{{\mathchoice {\setbox0=\hbox{$\displaystyle\rm Q$}\hbox{\raise
0.15\ht0\hbox to0pt{\kern0.4\wd0\vrule height0.8\ht0\hss}\box0}}
{\setbox0=\hbox{$\textstyle\rm Q$}\hbox{\raise
0.15\ht0\hbox to0pt{\kern0.4\wd0\vrule height0.8\ht0\hss}\box0}}
{\setbox0=\hbox{$\scriptstyle\rm Q$}\hbox{\raise
0.15\ht0\hbox to0pt{\kern0.4\wd0\vrule height0.7\ht0\hss}\box0}}
{\setbox0=\hbox{$\scriptscriptstyle\rm Q$}\hbox{\raise
0.15\ht0\hbox to0pt{\kern0.4\wd0\vrule height0.7\ht0\hss}\box0}}}}
\def\bbbt{{\mathchoice {\setbox0=\hbox{$\displaystyle\rm
T$}\hbox{\hbox to0pt{\kern0.3\wd0\vrule height0.9\ht0\hss}\box0}}
{\setbox0=\hbox{$\textstyle\rm T$}\hbox{\hbox
to0pt{\kern0.3\wd0\vrule height0.9\ht0\hss}\box0}}
{\setbox0=\hbox{$\scriptstyle\rm T$}\hbox{\hbox
to0pt{\kern0.3\wd0\vrule height0.9\ht0\hss}\box0}}
{\setbox0=\hbox{$\scriptscriptstyle\rm T$}\hbox{\hbox
to0pt{\kern0.3\wd0\vrule height0.9\ht0\hss}\box0}}}}
\def\bbbs{{\mathchoice
{\setbox0=\hbox{$\displaystyle     \rm S$}\hbox{\raise0.5\ht0\hbox
to0pt{\kern0.35\wd0\vrule height0.45\ht0\hss}\hbox
to0pt{\kern0.55\wd0\vrule height0.5\ht0\hss}\box0}}
{\setbox0=\hbox{$\textstyle        \rm S$}\hbox{\raise0.5\ht0\hbox
to0pt{\kern0.35\wd0\vrule height0.45\ht0\hss}\hbox
to0pt{\kern0.55\wd0\vrule height0.5\ht0\hss}\box0}}
{\setbox0=\hbox{$\scriptstyle      \rm S$}\hbox{\raise0.5\ht0\hbox
to0pt{\kern0.35\wd0\vrule height0.45\ht0\hss}\raise0.05\ht0\hbox
to0pt{\kern0.5\wd0\vrule height0.45\ht0\hss}\box0}}
{\setbox0=\hbox{$\scriptscriptstyle\rm S$}\hbox{\raise0.5\ht0\hbox
to0pt{\kern0.4\wd0\vrule height0.45\ht0\hss}\raise0.05\ht0\hbox
to0pt{\kern0.55\wd0\vrule height0.45\ht0\hss}\box0}}}}
\def\bbbz{{\mathchoice {\hbox{$\sans\textstyle Z\kern-0.4em Z$}}
{\hbox{$\sans\textstyle Z\kern-0.4em Z$}}
{\hbox{$\sans\scriptstyle Z\kern-0.3em Z$}}
{\hbox{$\sans\scriptscriptstyle Z\kern-0.2em Z$}}}}
\def\qed{\ifmmode\sq\else{\unskip\nobreak\hfil
\penalty50\hskip1em\null\nobreak\hfil\sq
\parfillskip=0pt\finalhyphendemerits=0\endgraf}\fi}
%

\magnification=\magstep1
\hsize=5.5 true in
\centerline{\bf NAKED SINGULARITIES IN SPHERICALLY SYMMETRIC INHOMOGENEOUS }
\centerline{\bf TOLMAN-BONDI DUST CLOUD COLLAPSE }
\vfill
\centerline{\bf P. S. Joshi and I. H. Dwivedi$^{*}$ }
\centerline{\bf Tata Institute of Fundamental Research}
\centerline{\bf Homi Bhabha Road, Bombay 400 005}
\centerline{\bf India}
\vfill
\noindent {\bf $^{*}$ Permanent Address}:\hfill\break
\noindent{\bf Institute of Basic Sciences}\hfill\break
\noindent{\bf Agra University}\hfill\break
\noindent{\bf Khandari, Agra, India}\hfill\break
\vfill
\noindent{\bf Proofs to be sent to:}\hfill\break
\noindent{\bf Dr. Pankaj  S. Joshi}\hfill\break
\noindent{\bf Theoretical Astrophysics Group}\hfill\break
\noindent{\bf T.I.F.R. Homi Bhabha Road}\hfill\break
\noindent{\bf Colaba, Bombay 400 005}\hfill\break
\noindent{\bf India}\hfill\break
\vfill

\vfil\eject

\magnification=\magstep1  
\hoffset=0 true cm        
\hsize=6 true in        
\vsize=8.5 true in        
\baselineskip=24 true pt plus 0.1 pt minus 0.1 pt 
   e spacing) 72.27 pt=1 inch
\overfullrule=0pt         

\centerline {ABSTRACT}

We investigate here the occurrence and nature of naked singularity
for the inhomogeneous gravitational collapse of Tolman-Bondi dust clouds.
It is shown that the naked singularities form at the center of the
collapsing cloud in a wide class of collapse models which includes the
earlier cases considered by Eardley and Smarr [5] and Christodoulou [6].
This class also contains self-similar as well as non-self-similar models.
The structure and strength of this singularity is examined and the question
is investigated as to when a non-zero measure set of non-spacelike
trajectories could be emitted from the singularity as opposed to isolated
trajectories coming out. It is seen that the weak energy condition and
positivity of energy density ensures that the families of non-spacelike
trajectories come out of the singularity. The curvature strength of the
naked singularity is examined which provides an important test for its
physical significance. This is done in terms of the strong curvature
condition which ensures that all the volume forms must be crushed to zero
size in the limit of approach to the singularity; and also the divergence
of the Kretschmann scalar ${\cal K}=R^{abcd}R_{abcd}$ is pointed out.
We show that the class considered here contains sub-classes of solutions
which admit strong curvature naked singularities in either of the senses
stated above. The conditions are discussed for the naked singularity to be
globally naked.  An implication for the fundamental issue of the final fate
of gravitational collapse is that naked singularities need not be considered
as artifacts of geometric symmetries of space-time such as self-similarity,
but arise in a wide range of gravitational collapse scenarios once the
inhomogeneities in the matter distribution are taken into account. It is
argued that a physical formulation for the cosmic censorship may be evolved
which avoids the features above. Possibilities in this direction are
suggested while indicating that the analysis presented here should be
useful for any possible rigorous formulation of the cosmic censorship
hypothesis.

\vfil\eject

\beginsection 1. INTRODUCTION

It is generally believed that a generic gravitational collapse would
commence from a highly inhomogeneous initial state.
This will be described in terms of an inhomogeneous energy density
distribution in terms of a regular
initial data on a spacelike hypersurface. The general
class of solutions
of Einstein's field equations describing spherically symmetric dust clouds,
independent of the homogeneity assumption,
was given by Tolman [1], which was further developed and studied by
Bondi [2]. This class could be used to model the gravitational
collapse of matter
from general inhomogeneous initial conditions and one can study the
fundamentally important issue of the final fate of gravitational
collapse of a massive star which has exhausted its nuclear fuel within this
framework. The assumptions involved here are the vanishing pressure and
the spherical symmetry of the matter distribution
which is in the form of dust. One could argue that in the final stages
of collapse, the matter distibution would become almost spherically
symmetric, and that the pressures should play a minor role to justify
the dust approximation. From our view point  however, the main advantage
is that subject to these conditions these models allow us to describe
the evolution of inhomogeneous
distributions of matter, which offers a very general class for the study
of the gravitational collapse phenomena.

A special case of the these Tolman-Bondi class of solutions is the
Oppenheimer-Snyder [3] study of a completely homogeneous dust cloud collapse
with zero pressure. This example has been studied in great detail and has
provided much insight towards
understanding
the final fate of a continually collapsing massive body such as a
star, which
could achieve no equilibrium state because of the dominance of gravitational
forces. This case provides the basic motivation for the idea of formation
of black holes as the final state of collapse,
and the related cosmic censorship hypothesis [4] which broadly states that the
singularities forming in gravitational collapse must  necessarily  be
hidden behind
the event horizons of gravity and hence permanently
invisible to the outside observers.
This cosmic censorship hypothesis plays a fundamental role in both the
theory and applications of the black hole physics and has been recognized
as one of the most important open problems in the general theory of relativity
and gravitation physics to-day.

As it turns out, despite several attempts no proof or any precise
mathematical formulation
of the cosmic censorship has been available so far.  Further, the
completely homogeneous dust collapse mentioned above
could also be viewed as a special case which forms a set
of zero measure in the general inhomogeneous class represented by the
Tolman-Bondi solutions. It thus becomes imperative to study the general
class of Tolman-Bondi models in greater detail in order to understand the
final fate of a gravitationally collapsing massive body when the
effects of inhomogeneities are taken into account. In fact, it was
pointed out by the numerical simulations of
Eardley and Smarr [5] that naked singularities not covered
by event horizons arise in the marginally bound Tolman-Bondi collapse, and
subsequently  a class of such models was studied in detail analytically
by Christodoulou [6] to
draw the same conclusion. However, these singularities were shown to be
gravitationally weak by Newman [7], who studied the curvature strengths of
such naked singularities and conjectured that nature avoids strong
curvature naked singularities.

Our purpose here is to study the Tolman-Bondi inhomogeneous collapse
for a rather general class of models, which includes the above classes,
and to study the formation and structure of the naked singularity occurring
at the center of the collapsing cloud.
We show that the formation of  naked singularity is a generic feature
for a very wide range of solutions considered
here. We have recently shown [8] such a result for the general class of
self-similar models describing the gravitational collapse of a perfect fluid,
where it is shown that a powerfully strong curvature naked singularity
forms from which families of non-spacelike geodesics escape in the space-time.
Further to this, the class considered here is shown to include all the
self-similar Tolman-Bondi models as well as a wide range of non-self-similar
models. This indicates that the naked singularity may not be regarded as
the consequence of the geometric property of self-similarity only [9].
A naked singularity may not be
treated as a serious enough situation if only a single non-spacelike
trajectory escaped from it. Thus, we examine the sufficient conditions
when families
of non-spacelike geodesics could escape from the naked singularity.
Interestingly, it turns out that the validity of weak energy condition
in the space-time ensures the existence of such families. This is analogous
to the  results of Ref.8 for the self-similar class. We also
discuss the issue as to when the naked singularity will be globally naked,
i.e. visible to far away observers.

The organization of the paper is as below.
In section 2, the basic parameters of the Tolman-Bondi models
describing the inhomogeneous dust collapse are specified. The existence and
structure of the naked singularity is analyzed in section 3. We
also characterize here the conditions that ensure
that families of non-spacelike
geodesics, rather than a single isolated trajectory,
are emitted from the naked
singularity. In particular, it is shown that the weak energy condition,
together with the positivity of energy density,
implies that a non-zero measure set of non-spacelike geodesics comes out
from the naked singularity. The global versus local nakedness
of the singularity is also discussed here.
The curvature strength of the naked
singularity provides an important test of the physical significance for the
same. This issue is examined in section 4, where it is shown
that the models considered
here include both self-similar as well as non-self-similar classes
admitting a strong curvature singularity in a powerful sense. The final
section 5 summarizes some of the implications and conclusions.

\vfil\eject

\beginsection 2. TOLMAN-BONDI SPACE-TIMES

The Tolman-Bondi metric representing collapse of a spherically symmetric
inhomogeneous dust cloud in the comoving coordinates (i.e. $u^i=\delta^i_t$)
is given by
$$ds^2= -dt^2+{R'^2\over1+f}dr^2+R^2(d\theta^2+sin^2\theta d\phi^2)\eqn\qq$$

$$T^{ij}=\epsilon \delta^i_t \delta^j_t,\quad \epsilon=\epsilon(t,r)={F'
\over R^2R'}\eqn\qq$$
where $T^{ij}$ is the stress-energy tensor, $\epsilon$ is
the energy density and $R$ is a function of both $t$ and $r$ given by
$$\dot R^2={F\over R}+f\eqn\qq$$
Here the dot and the prime denote partial derivatives with respect
to the parameters $t$ and $r$ respectively
and as we are only concerned with the gravitational collapse
of dust, we require $\dot R(t,r)<0$.
The quantities
$F$ and $f$ are arbitrary functions
of $r$. The quantity $4\pi R^2(t,r)$ gives the proper area of the  mass shells
and the area of such a shell at $r=const.$ goes to zero when $R(t,r)=0$.
Integration of equation (3) gives
$$t-t_0(r)=-{R^{3/2}G(-fR/F)\over \sqrt{F}}\eqn\qq$$
where $G(y)$ is a strictly real positive and bound function which has the
range $1\ge y\ge - \infty$ and is given by
$$\eqalign{ G(y)&=\left ( {\sin^{-1}\sqrt{y}\over y^{3/2}}-{\sqrt{1-y}\over y}
\right )\quad\hbox{for}\quad 1\ge y>0 \cr
G(y)&={2\over3}\quad \hbox{for}\quad y=0\cr
G(y)&=\left ({-\sinh^{-1}\sqrt{-y}\over (-y)^{3/2}}-{\sqrt{1-y}\over y}
\right )\quad\hbox{for}\quad 0> y\ge -\infty \cr}\eqn\qq$$
and $t_0(r)$ is a constant of integration. We thus have in
all three arbitrary functions of $r$ namely $f(r)$, $F(r)$ and $t_0(r)$.
One could however, use
the remaining  coordinate freedom left in the choice of scaling of $r$
in order to reduce the number of such arbitrary functions to two.
We therefore rescale $R$ using  this coordinate freedom
such that
$$R(0,r)=r\eqn\qq$$
Then $t_0(r)$ is evaluated by using the equation above and (4) to give
$$t_0(r)={r^{3/2}G(-fr/F)\over \sqrt{F}}\eqn\qq$$
The time $t=t_0(r)$ corresponds to the value
$R=0$ where the area of the shell of
matter at a constant value of the coordinate $r$ vanishes. It follows that
the singularity curve $t=t_0(r)$ corresponds to the time when the matter shells
meet the physical singularity. Thus, the range of the coordinates is given by
$$ 0\le r<\infty, \qquad -\infty<t<t_0(r)\eqn\qq$$
It follows that unlike the collapsing
Friedmann case, where the physical singularity
occurs at a constant epoch of time (say, at $t=0$), the singular epoch now
is a function of $r$ as a result of inhomogeneity in the matter distribution.
One could recover the Friedmann case from the above equations if we set
$t_0(r)=t'_0(r)=0$.

The function $f(r)$ classifies the space-time as bound, marginally bound, or
unbound depending on the range of its values
which are
$$f(r)<0, \quad f(r)=0,\quad \hbox{and} f(r)>0$$
respectively.
The function $F(r)$ can be interpreted as the weighted mass
(weighted by the factor $\sqrt{1+f}$)
within the dust ball ${\cal B}$ of coordinate radius $r$
which is conserved in the following
sense.
$$m(r)={F(r)\over 2}
=\int_{\cal B}(1+f)^{1/2}\epsilon (t,r)dv = 4\pi\int_0^r
\rho(r)r^2dr\eqn\qq$$
where $\epsilon(0,r)=\rho (r)$.
For physical reasonableness the weak energy condition would be assumed
throughout the space-time, i.e. $T_{ij}V^iV^j\ge 0$ for all non-spacelike
vectors $V^i$. This implies that the energy density
$\epsilon$ is everywhere positive, ($\epsilon \ge 0$)
including the region near  $r=0$.
Partial derivatives
of $R$ like $R'$ and $\dot {R'}$ are of importance in our analysis.
We get
from the equations (3) to (7)
$$R'=r^{\alpha -1}\left((\eta-\beta)
X+(\Theta-(\eta-{3\over 2}\beta)
X^{{3\over 2}}G(-PX))(P+{1
\over X})^{1\over 2}\right)\equiv r^{\alpha-1}H(X,r)\eqn\qq$$
$$\dot R'=
{\Lambda ^{1\over 2}\over 2rX^2}\left(-\beta X^2({1\over X}+P)^{{1\over 2}}
 +\Theta
-(\eta-{3\over 2}\beta)
X^{{3\over 2}}G(-PX)\right)\equiv {-N(X,r)\over r}\eqn\qq$$
where we have put
$$X=(R/r^{\alpha}),\quad \eta=\eta(r)=r{F'\over F}, \quad \beta=\beta(r)=
r{f'\over f},\quad p=p(r)=rf/F\eqn\qq$$

$$P=pr^{\alpha-1},\quad \Lambda={F
\over r^{\alpha }},\quad \Theta\equiv {t_0'\sqrt{\Lambda}\over r^{\alpha-1}}
={1+\beta-\eta \over
(1+p)^{1/2}r^{3(\alpha -1)/2}}+{(\eta -{3\over 2}\beta)G(-p) \over
r^{3(\alpha -1)/2}}\eqn\qq$$
The function $\beta(r)$ is defined to be zero when $f$ is constant and zero.
The factor  $r^{\alpha }$ has been introduced here for the sake of
convenience in examining the structure of the naked singularity.
The exact value of the
positive constant $\alpha \ge 1$ is to be determined and will depend on the
different models of the space-time which allow naked singularities. Functions
$H(X,r)$ and $N(X,r)$ are defined by equations (10) and (11). Using the
scaling given by (6), the energy density $\epsilon$ on the hypersurface
$t=0$ is written as $\epsilon=F'/r^2$. Since the weak energy conditions are
satisfied and $F$ is a function of $r$ only, it follows that $F'\ge0$
through out the space-time. One can write the energy density as
$$\epsilon = {\eta \Lambda \over R^2 H}\eqn\qq$$
Since $F'=\eta\Lambda r^{\alpha-1}$, it follows from the above that
everywhere  $H(X,r)\ge 0$ and $\eta \Lambda \ge 0$ everywhere as a consequence
of the weak energy condition.

Singularities are the boundary points of  the space-time where the normal
differentiability and manifold structures
break down. In other words, these are the
points where the energy density given by equation
(2), or the curvature quantities such as the scalar polynomials constructed
out of the metric tensor and the Riemann tensor diverge. One example of
such a quantity is the Kretschmann scalar ${\cal K}=R_{abcd}R^{abcd}$, which
is given in the Tolman-Bondi case by
$${\cal K}=12{F'^2\over R^4R'^2}-32{FF'\over R^5R'}+
48{F^2\over R^6}\eqn\qq$$
Such singularities are indicated by the existence of
incomplete future or past directed non-spacelike geodesics in the space-time
which terminate at the singularity. Then one requires that the curvature
quantities stated above assume unboundedly large values in the limit of
approach to the singularity along the non-spacelike geodesics terminating
there. If such a condition is satisfied, then one would like to consider
the singularity to be a physically significant curvature singularity.

In Tolman-Bondi space-times singularities occur, as
one can see from equations (2) and (15), at points where  $R=0$,  which
are called shell focusing singularities, and also at points where $R'=0$.
At the points where $R'=0$ the Tolman-Bondi metric is
degenerate and these are called shell crossings.
In the context of Tolman-Bondi space-times the points
$R>0,F'>0 $, where $R'=0$, are called the
shell crossing singularities [7].
Such shell crossing singularities
in Tolman-Bondi space-times
have been analyzed in detail in the literature [10,11], and their nature
appears to be fairly well understood.
Even though such shell-crossing singularities  could be locally
naked, the important point is
they have been shown to be gravitationally weak [7]. Thus it is
generally believed that such shell crossing singularities
need not be taken seriously as far as the cosmic
censorship conjecture is concerned. The absence of shell-crossing
singularities in a space-time turns out
to be related to the condition that the function $t_0(r)$ giving the proper
time for the shells to fall into the physical singularity should be a
monotonically increasing function.
The dust density and certain components
of the curvature blow up near  such a singularity. However,
the causal structure of the space-time can be extended through
such a singularity and the space-time metric also can be defined
in the neighborhood of such a point in a distributional sense [12].
In the context of such a situation, we do not consider here such shell
crossings, and assume that  there are
no shell crossing singularities in the space-time (except probably
right at the center $r=0$ [10]).
This does not involve any loss of generality as our basic purpose here is
to examine the formation and local structure of the shell focusing
naked singularity at the center of the collapsing cloud.
Whereas the existence
of shell crossings will not affect the qualitative nature of these general
conclusions, the above assumption
allows the calculations to be presented in a more transparent manner.

Unlike the shell crossings,
the space-time metric however, admits no extension through a
shell focusing singularity occurring at $R=0$
which is more difficult to ignore. A shell-focusing singularity
can be avoided only by rejecting
the forms of matter such as dust as the fundamental forms of matter
( see e.g. [5]).
Hence, we investigate here the occurrence of such
shell focusing singularities at he center of the collapsing dust
cloud and examine their nature and structure
for the Tolman-Bondi space-times.
It has been shown earlier [6] that a shell focusing singularity
occurring at $r>0, R=0$ is totally spacelike and therefore our discussion would
be confined to the singularity at $r=0$.

The points $(t_0,r_0)$
where a shell focusing singularity $R(t_0,r_0)=0$ occurs,
are related by equation (4). The singularity here
occurs at $r=r_0$ at the coordinate time $t=t_0$ and we would
call the singularity to be a {\it central singularity} if it occurs at $r=0$.
Earlier work [3,4]
has shown that this central shell focusing
singularity is naked, though gravitationally weak, for a class of Tolman-Bondi
space-time for which the energy density (which is assumed to be
positive every where
and is taken to be non-zero at $r=0$) and the metric are
even smooth functions of $t$ and $r$.
Translated in terms of parameters defined above,
this corresponds to the  class for which $\eta(0)=3,\beta(0)=2$,
and $p(r)$ is an
even smooth function of $r$.
In terms of functions $F(r)$ and $f(r)$ it amounts to the conditions,
$$F(r)=r^3{\cal F}(r),\qquad \infty >{\cal F}(0)>0,\qquad 0<p(r)\le 1\eqn\qq$$
It was however, pointed out by Waugh and Lake[13] and Ori and Piran[14]
that this class of
gravitationally weak naked singularities excludes the self-similar Tolman-Bondi
models, where they showed the singularity to be gravitationally strong
along the Cauchy horizon, which is a null geodesic coming out of the
singularity. Further, Grillo [15] pointed out an example in the case
of unbound Tolman-Bondi
models which are non-self-similar and the naked singularity is gravitionally
strong. In fact, we have pointed out recently [16] that the naked singularity
exists and is gravitationally strong for a wide class of Tolman-Bondi models
which are non-self-similar in general and include all the self-similar
models as a special subclass.
In the notation used here, these models are characterized by the conditions
$\eta(0)=1$ with $F(r)$ and $f(r)$ being analytic at $r=0$.

Through out the present consideration we would require
rather general differentiability  conditions on
the functions $F(r)$ and $f(r)$ in that they will be assumed to be
atleast $C^1$
at the center $r=0$, $\infty>\eta (0)>0$, and $\beta(0)$ is finite [17].
We note that the function $f$, and also its first
derivatives (through $R'$) enter the metric potentials.
One might actually argue that the above is a more general condition
than should be required, because it is often a customary practice to
assume that the metric is $C^2$-differentiable
(which ensures again the above requirement),
so that the metric transformations and other functions
connected with the metric are well-defined to do regular physics.
Hence, such a condition may be considered to be physically reasonable
and a rather general differentiability requirement which
includes practically all the inhomogeneous collapse
Tolman-Bondi models of interest. In fact, one could argue that if the
metric is not $C^2$-differentiable, but say only $C^1$ on initial
surface, it may be considered as being already
naked singular and not defining a regular initial data on an initial
spacelike hypersurface.

In order to represent the gravitational collapse scenario,
we assume the energy density $\epsilon$ to have a compact support
on an initial spacelike hypersurface and the Tolman-Bondi space-times
given by (1) can be matched at some $r=const=r_c$
to the exterior Schwarzschild field
$$ds^2=-(1-{2M\over r_s})dT^2+{dr_s^2\over 1-{2M\over r_s}}+r_s^2d\Omega ^2
\eqn\qq$$
where $d\Omega^2=d\theta^2+sin^2\theta d\phi^2$. The value of the
Schwarzschild radial coordinate is $r_s=R(t,r_c) $ at the boundary $r=r_c$.
We have $m(r_c)=M$ where $M$ is the
total Schwarzschild mass enclosed within the dust ball of coordinate radius of
$r=r_c$. Without going into further
details of the matching conditions we would like to say a few words regarding
the apparent horizon. The apparent horizon in the interior dust ball
lies at $R=F(r)$. From (4) and (7) one can see that the corresponding time
$t=t_{ah}(r)$ is given by
$$t=t_{ah}(r)={r^{3/2}G(-p)\over \sqrt{F}}-FG(-f)\eqn\qq$$
It has been shown earlier [6,7] that emissions from the shell focusing
singularity $R(t_0,r_0)=0$ for all $r_0>0$
would lie in the region above $t=t_{ah}$ i.e. $t_0>t_{ah}$ for all $r_0>0$,
$t_0$
being the time when singularity at $r=r_0$ occurs. Hence all radiations
would be future trapped from shell focusing singularities at $r>0$.
At $r=0$ however, $t(0)=t_{ah}(0)$ and the singularity could be atleast
locally naked. Any light ray terminating at this singularity in the
past goes to the future infinity if it reaches the surface of the
cloud $r=r_c$ earlier than the apparent horizon at $r=r_c$. In
such a case the singularity would be globally naked.
We now examine this central singularity in the section below.

\vfil\eject

\beginsection 3. THE EXISTENCE AND STRUCTURE OF NAKED SINGULARITY

In this section we  investigate the existence of the naked
central singularity for the general class of Tolman-Bondi space-times
under consideration. The singularity is naked if there are future
directed non-spacelike curves in the space-time with their past
end-point at the singularity. The existence of such curves implies
that either photons or timelike particles can be emitted from the
singularity. In particular we will examine the
future directed  non-spacelike geodesics for their past end-point
at the singularity. Other related issues examined here are when a non-zero
measure set of non-spacelike trajectories will meet the singularity in the
past, rather than a single isolated geodesic; and when such a singularity
will be globally naked.

\centerline {A. The Existence}

The tangents $K^a=dx^a/dk$ for the outgoing
non-spacelike geodesics in the Tolman-Bondi space-time given by (1)
can be written as below,
$$K^t={dt\over dk}={{\cal P}\over R}\eqn\qq$$
$$K^r={dr\over dk}={\sqrt{1+f}\sqrt{{\cal P}^2-\ell^2+BR^2}\over RR'}\eqn\qq$$
$$(K^{\theta})^2+sin^2
\theta (K^\phi)^2=\ell^2/R^4\eqn\qq$$
Here  $\ell$ is an impact parameter that labels different geodesics and
vanishes($\ell =0$) for radial trajectories, $B$ characterizes the
type of geodesics i.e.
$B=0$ for null and $B=-1$ for timelike curves,
and the function ${\cal P}=
{\cal P}(t,r)$ satisfies the differential equation
$${d{\cal P}\over dk}+({\cal P}^2-\ell^2+BR^2)
({\sqrt{1+f}\dot R'\over RR'}-{\dot R
\over R^2})-({\cal P}^2-\ell^2+BR^2)^{1/2}{\cal P}{\sqrt{1+f}\over R}
+B\dot R=0\eqn\qq$$
The parameter $k$ is an affine parameter along the geodesics.
For future directed non-radial trajectories
that meet the central singularity at $R=0$ in past, it follows from
equation (20) that ${\cal P}\ge \ell$ near the singularity.

If the outgoing non-spacelike geodesics are to terminate in the past
at the central singularity
$r=0$, which occurs at some time $t=t_0$ at which $R(t_0,0)=0$, then
along such
geodesics we have $R\rightarrow 0$ as $r\rightarrow 0$.
The following is satisfied along non-spacelike geodesics

$${dR\over du}={1\over \alpha r^{\alpha -1}}
(\dot R{dt\over dr}+ R'
)=\left(1-{{\cal P}\sqrt{f+{\Lambda \over X}}
\over \sqrt{1+f}\sqrt{{\cal P}^2-\ell^2
+BR^2}}\right){H(X,u)\over \alpha }\equiv U(X,u)\eqn\qq$$
where we have put $u=r^{\alpha}$.
The function $H(X,r)$ in the above equation is strictly positive and
non-zero for all $r>0$ as a consequence of (10) and (14).
For an outgoing geodesic $(dR/du)$ must be positive while the negative value
for this quantity means the geodesic is ingoing.
The point $u=r^{\alpha}=0,R=0$ is a
singularity of the above differential equation.

It is now essential
to understand the exact nature of this singularity.
If the functions appearing in the numerator and denominator
of (23) are expandable and contain linear terms, then one can apply
the standard analysis on the classification of singular points
of first order differential equations [18] to understand the nature
of this singularity. However, in the case otherwise, the same could be
understood only by means of studying
the detailed behavior of the characteristic curves in the
vicinity of the singularity.
If these characteristics terminate at the singularity in past
with a definite tangent, this is determined  by the limiting
value of $X=R/r^{\alpha}=R/u$ at $R=0,u=0$.
If the non-spacelike geodesics meet the singularity
with a definite value of tangent, then using equation (23) and l'Hospital rule
we get for the value of $X_0$,
$$X_0=\lim_{R\rightarrow 0,u\rightarrow 0}
{R\over u}=\lim_{R\rightarrow 0,u\rightarrow 0}{dR\over du
}=
\lim_{R\rightarrow 0,u\rightarrow 0}U(X,u)=U(X_0,0)\eqn\qq$$
If a real and positive value of $X_0$ satisfies the above
equation then the singularity
could be naked.
Real and positive roots of the above equation gives the possible
values of tangents the  outgoing geodesics can have at the singularity.
Thus, if a real and positive value of $X=X_0$ satisfying above equation
exists, then integral curves of the differential equation (23)
i.e. outgoing non-spacelike geodesics, can terminate in the past
at the singularity
with
a definite value of the tangent given by $X=X_0$.
Clearly if no real positive root of the above exists then
the singularity is not naked.

In order to make the discussion transparent, at this point
we would limit ourselves to radial null geodesics only.
Similar consideration can be given for the non-radial non-spacelike
geodesics as well in terms of (23) and (24),
which will be more complicated in view
of the assumed generality of the functions involved. The equations
(23) and (24) could then be written as
$${dR\over du}=\left(1-{\sqrt{f+{\Lambda \over X}}
\over \sqrt{1+f}
}\right){H(X,u)\over \alpha }\equiv U(X,u)\eqn\qq$$
$$V(X_0)=0\eqn\qq$$
where
$$V(X)= U(X,0)-X=
\left(1-{\sqrt {f_0+{\Lambda_0
\over X}}\over \sqrt{1+f_0}}\right){H(X,0)\over \alpha}-X\eqn\qq$$
where we have introduced the notation
$$\beta_0=\beta(0),
\eta_0=\eta(0), f_0=f(0),
P_0=P(0), \quad \Lambda_0=\Lambda(0),\quad \Theta_0=\Theta (0)
\eqn\qq$$
Along an outgoing null geodesic from the singularity, $r$ increases and
so does the area coordinate $R$.
A point to note is that $dR/du$ is
positive for $X>\Lambda$, which implies $R>F$, and the geodesic
is outgoing. If the
geodesics cross and get inside the curve $R=F$, which represents the
apparent horizon, $dR/du$ becomes negative and hence
the geodesics are ingoing
(in the sense that area coordinate $R$ starts decreasing).
Since the apparent horizon $R=F$ is the boundary of all trapped surfaces,
if a null geodesic terminating at the singularity is to be outgoing
it must have $R>F$ at the singularity along the geodesic. The null geodesic
would also reach the future infinity if it does not get inside the apparent
horizon(i.e. $R<F$) within the boundary of the dust cloud and reaches this
boundary at $r=r_c$ with $R>F$ along the same geodesic.

In the description given here the constant $\alpha$ actually
represents the behavior of
singular geodesics near the singularity i.e. $R\propto r^{\alpha}$ near
the singularity. In fact we can write
$R=X_0r^{\alpha}$ in the neighborhood of singularity,
$X_0$ being the real and positive root of equation (26).
Thus the determination
of $\alpha$ really means determining the behavior of possible
singular geodesics terminating at the singularity.
The alogrithem for  evaluation of the value of $\alpha$ is as follows:
Given the functions $F(r)$ and $f(r)$ (which specify the Tolman-Bondi model),
the unique value of $\alpha$ is  determined by the condition that
$\Theta(r)\sqrt{P+{1\over X}}$
does not vanish or goes to infinity identically as $r\to 0$ in the limit of
approach to the central singularity along any $X=const$ direction.
This condition ensures that the quantity
$H(X,0)$ will not be identically zero or infinite  in (26) and (27).
Note that $\Theta(r)$ vanishes
identically only for the case of
Friedmann models ($\eta(r)=3, \beta(r)=2$) where the
singularity is spacelike.
Once such a value of $\alpha$ is determined the  values of
positive roots of the equation (22) are then determined if there are any.
There remains a possibility when such a value of $\alpha$ can not be
found. Such a case can arise only in the some of the situations
where $\beta(0)=2, \eta (0)>3$.
In this case, actually one has $\Theta (r) \propto (r^{\eta_0-2-\alpha}\ln r)$
near the singularity at $r=0$. However, in this situation one can use a
suitable change of the variable $R$, namely,
$\bar R=R+ar^{\eta_0-2}(\ln r +b)$ and
$\bar X=X+ar^{\eta_0-2-\alpha}(\ln r +b)$,
($a$ and $b$ are some constants). This  reduces  equation (25) and (26) in the
desired form and the value of $\alpha$ can then again be determined.
Once the value of $\alpha$ is known in this manner,
one can easily establish whether
the singularity could possibly be naked. That is, if
for this value of $\alpha$ the quantity $\Lambda_0$ diverges then clearly
the space-time does not permit a naked singularity as $X_0=-\infty$.
In fact, this puts an upper bound on the possible values
of $\alpha$ if the singularity is to be naked, which is
given by $\alpha \le \eta_0$. It follows from equation (26) that $V(0)\neq 0$
hence $X=0$ can not be the root of $V(X)=0$ and this implies
that $H_0=H(X_0,0)
\neq 0$.
We specify  the values of $\alpha$
for some specific classes. If for example,
$\eta_0
=1$ then $\alpha =1$.
It should be noted that for the case when $\eta_0=3, \beta_0=2$ (the cases that
have been discussed in [5] and [6,7]),
and $F$ and $f$ are even functions of $r$, the value of $\alpha$ turns out to
be
$$\alpha =7/3 \quad X_0=({3\over 4}\Theta_0)^{2\over 3}\eqn\qq$$
In these cases a shell crossing singularity
also occurs at the central singularity (i.e. $R'=0$) along with the
shell focusing singularity. Again, $\alpha$ determines the occurrence of a
shell crossing singularity at the central singularity.
It follows from equations (10) and (26)
that near the central singularity at $r=0$, $R'=r^{\alpha-1}H(X_0,0)$.
Hence, for $\alpha >1$ the shell crossing singularity would also occur along
with a shell focusing one. This actually happens
in the cases already discussed by
the references [5] and [6]. On the other hand, if $\alpha=1$ no shell
crossing singularity occurs at the central singularity
as the cases discussed in Ref.15.

This determination of the value of $\alpha$ allows one to determine
the existence of real and positive roots of equation (26).
If the equation $V(X)=0$ has a real and positive root, the singularity
could be naked
and the geodesics could terminate at the
singularity in past with the tangent $X=X_0$ in the $(u,R)$ plane.
Therefore, existence of atleast one
real positive root of
(26) is the necessary condition for the space-time to admit
naked singularity. The positive root $X=X_0$ actually represents the
value of the tangent to null geodesics at the singularity, and it follows
from equation (26) that $X_0>\Lambda _0$. Since $\Lambda_0$ is
the value of the tangent of the apparent horizon $R=F$ [19]
at the singularity, it
is clear that the geodesics in such cases could be at least locally naked.
Clearly if no real positive root of the above is found, the
singularity $R=0,r=0$ is not naked.
It should be noted that
many real positive roots of the equation (26) may exist which give the
possible values of tangents to the null geodesics
at the singularity. It is possible however, that the
integral curves may or may not realize a given
value $X_0$ at the singularity.

To determine whether a value $X_0$
is realized at the naked singularity along any outgoing singular geodesic,
which establishes the nakedness of the singularity,
consider the
equation of radial null geodesics in the form $u=r^{\alpha}=u
(X)$. From equation
(25) we have
$${dX\over du}={1\over u}({dR\over du}-X)=
{U(X,u)-X
\over u}\eqn\qq$$
The solution of the above gives trajectories
of radial null geodesics in the form $u=u(X)$.
The necessary condition for a null geodesic to terminate at the
singularity at $R=0,u=0$ is that
$V(X)=0$ must have a real positive root $X=X_0$. In such a case,
non-spacelike curves could terminate
at the singularity with the tangent $X_0$. Therefore, if the null geodesics
do terminate at the singularity then $u\rightarrow 0$ as $X\rightarrow X_0$
along the same. Let $X=X_0$ be a simple root of the equation (26).
We could then write
$$V(X)\equiv (X-X_0)(h_0-1)+ h(X)\eqn\qq$$
where $h_0$ is a constant the value of which is determined in terms of the
quantities defined earlier as,
$$h_0={1\over H_0}\left({\Lambda_0
H_0^2\over 2\alpha X_0^2\sqrt{f_0+{\Lambda_0\over X_0}}\sqrt{f_0+1}}+{X_0
N_0\over \sqrt{f_0+{\Lambda_0\over X_0}}}\right)\eqn\qq$$
The function  $h(X)$ is so chosen that
$$h(X_0)=\left({dh\over dX}\right)_{X=X_0}=0$$
i.e. $h(X)$ contains higher order terms
in $(X-X_0)$ and $H_0=H(X_0,0), N_0=N(X_0,0)$.
Equation (30) could then be written as
$${dX\over du}-(X-X_0){(h_0-1)\over u}={S\over u}\eqn\qq$$
where $S=S(X,u)=U(X,u)-U(X,0)+h(X)$
is such that $S(X_0,0)=0$, i.e. in the limit as
$u=0, X=X_0$ we have $S\to 0$.
Integration of (33) gives the equation of geodesics as $u=u(X)$.
Multiplying  (33) by $u^{-h_0+1}$ and integrating gives
$$ X-X_0= Du^{h_0-1} +u^{h_0-1}\int Su^{(-h_0+1)}du\eqn\qq$$
where $D$ is a constant of integration that labels different geodesics.
If the singularity is the end point of these geodesics with tangent
$X=X_0$, we must have $X\to X_0$ as $u\to 0$ in (34). Note that
as  $X\to X_0, u\to 0$, the last term in equation (34) always vanishes
near the singularity regardless of the value of the
constant $h_0$
(this is due to the reason that as
$u\to 0, X\to X_0$,  we have $S\to 0$). The first term on the
right-hand side of the equation, namely $Du^{h_0-1}$, however, vanishes only
if $h_0>1$.
It follows therefore that the single null geodesic described by $D=0$
always terminates at the singularity $R=0,u=0$, with $X=X_0$ as tangent.
On the other hand, if $h_0>1$ a family of outgoing singular geodesics
terminates
at the singularity with each curve being labeled by different values
of constant $D$.

Therefore, if a real and positive root of the equation  (26)
exists then singularity
will always be atleast locally naked. It follows that the existence of a real
and positive root of the equation (26) is both the necessary and sufficient
condition for the singularity to be locally naked.

The above analysis implies that a very wide class of Tolman-Bondi
space-times would in fact allow the existence of a naked singularity. In the
following we consider a few examples which illustrate this point and
provide insight into the formalism. Let us consider
first the marginally bound Tolman-Bondi space-times characterized by the
functions $F(r)$ and $f(r)$ as
$$f(r)=0,\quad F(r)=F_0 r^{n},\quad n\neq 3,n\ge 1
\eqn\qq$$
In the above $F_0$ is to be treated as a constant.
In this case, the relevant functions and equation (26) are as below.
$$ \alpha=1,\quad H(X,r)={nX\over 3}+{3-n\over 3\sqrt{X}},\quad
 \Lambda(r)=F_0r^{n-1}$$
$$V(X)=(3-n)X+n\sqrt{\Lambda(0)} \sqrt{X}-
{(3-n)\over \sqrt{X}}+{(3-n)\sqrt{\Lambda(0)}\over X}=0\eqn\qq$$
In the case $n>1$, where $\Lambda(0)=0$,
 the above equation has only one positive root
$X=1$ which satisfies the equation $V(X)=0$ for all $n>1$, thus
establishing the existence of naked singularity for all these space-times.
These results agree with the earlier numerical calculations of [5]
for the cases $n={9\over 5}$ and , $n={9\over 7}$.
In case $n=1$ the space-time is self-similar with $\Lambda(0)=F_0$ and the
equation (36) becomes
$$2x^4+x^3\sqrt{F_0}-2x+2\sqrt{F_0}=0\eqn\qq$$
where we have put $x^2=X$. The above has real and positive roots if
$$(F_0)^{{3\over 2}}< 4(26-15\sqrt{3})\eqn\qq$$
For example, for $\sqrt{F_0}=7/17$ there are two positive roots $x=0.5$ and
$x=0.658$. Hence for all such values given by equation (38) the singularity
is naked.

Next, consider the Tolman-Bondi space-times defined by the
values of $F$ and $f$ given by,
$$f(r)=f_0r^2(1+f_1r^3),\quad F(r)=F_0r^3,\quad {f_0\over F_0}=p_0>-1
\eqn\qq$$
Here $f_0, F_0$ and $f_1$ are to be treated as  some constants.
For this second example the relevant quantities are written as
$$\beta_0=2,\quad \eta(r)=3,\quad p(r)=p_0(1+f_1r^3)$$
$$\alpha =3,\quad
\Theta_0= f_1({1\over \sqrt{1+p_0}}-{3\over 2}G(-p_0)),\quad \Lambda(r)=F_0$$
$$V(X)=0\Rightarrow 2x^4+x^3\sqrt{F_0}-\Theta_0 x+\Theta_0\sqrt{F_0}=0
\eqn\qq$$
Where  we have again put $X=x^2$ and we see that for a wide range
of constants $f_0,F_0, f_1$ the positive root of the above would exist
and the singularity would be naked. In fact, for
$${\Theta_0\over (F_0)^{3/2}}>13+{15\over 2}\sqrt{3}\eqn\qq$$
the above equation always has two real positive roots establishing the
nakedness of the singularity.
These space-times are effectively of the type as those considered by
Newman [7], however, a condition on the evenness of functions was assumed
there which we have relaxed here.

\vfil\eject

\centerline {B. The Structure of Singularity}

We have shown above that if a real positive
root of $V(X)=0$ exists then atleast one single outgoing geodesic would
terminate at the singularity in the past and thus the singularity would
be naked.
If a single ray in the $(u,R)$ plane escapes from the singularity it
amounts to a single wavefront being emitted,
and thus the singularity appears naked only
instantaneously to a distant observer.
If the singularity is to be naked for a finite
period of time a non-zero measure set of null geodesics (i.e. families of null
geodesics) must have the singularity as their past end point. In earlier
examples of a naked singularity occurring in Vaidya space-times [20]
and in self-
similar space-times  families of non-spacelike geodesics terminate at the
naked singularity in the past.
In fact an analysis of self-similar gravitational
collapse of a perfect fluid in order to examine the nature and structure
of naked singularity has shown [8] that a non-zero measure
of non-spacelike geodesics terminate at the singularity in past provided
the weak energy condition and positivity of energy are
not violated in the near regions of
the singularity. This results into the exposure of the singularity to a distant
observer for an infinite period of time.
We therefore examine this issue of termination of families of non-spacelike
geodesics at the singularity below.

It follows from equation (34) that when only one simple real positive
root $X=X_0$ for $V(X)=0$ exists, no families of geodesics
would terminate at the singularity
if $h_0\le 0$. On the other hand, if $h_o>1$ it is  seen that
an infinity of
integral curves will meet the singularity in the past with tangent $X=X_o$,
different curves being characterized by different values of the constant $D$.
Thus, one sufficient condition for the families of non-spacelike curves
to meet the naked singularity in past is $h_0>1$, when $V(X)=0$ admits only
one simple real positive root. Such a condition corresponds to the
requirement that $h_0-1=(dV/dX)_{X=X_0}$ must be positive,
i.e. $V(X)$ must be an increasing
function at $X=X_0$. The interpretation of such a condition is seen very
clearly in the
case of self-similar models [13,8],
where this derivative of $V$ is determined
directly by the Einstein field equations in terms of the energy density and
the components of the metric tensor. It turns out in that case that this
derivative will be positive with $h_0>1$ provided the weak energy condition
is satisfied and the energy density is always greater than a certain lower
bound in the neighborhood of the singularity,
which gives a sufficient condition for families to meet the naked
singularity in the past.

Suppose now that equation (26) has two simple positive roots $X_1$ and $X_2$.
In such a case at least one singular geodesic would always terminate along
each of the tangents $X=X_1$ and $X=X_2$ at the singularity. Furthermore,
since $V(X)=0$ has two simple roots it follows that the value of its derivative
$h_0-1$ would be negative along one of the root and positive along
the other. Therefore, atleast along one of the roots $h_0>1$.
Hence the situation that emerges is that in such a case
families of geodesics will always terminate along one of the roots for which
$h_0>1$
while along the other only a single geodesic would escape.
The conclusions are the same if
$V(X)=0$ has more than two simple positive roots.
Thus existence of two positive roots is a sufficient condition for
a non-zero set of geodesics to terminate at the singularity.

This situation is similar to the scenario arising in the gravitational
collapse of radiation shells which we have analyzed  in detail
for the case of a linear mass function in Vaidya space-times [20], where
the full structure of families of all the
non-spacelike geodesics terminating at the naked singularity in the past
has been worked out.
It is seen there that when the corresponding quantity there has two roots,
they provide the tangent values for the escaping geodesics. The families
of non-spacelike geodesics
meet along one of the roots as the tangent at the naked singularity, where as
there is a single null trajectory escaping from the singularity at the
second root. In fact, Lemos [21] has pointed out recently several parallels
between the self-similar Tolman-Bondi models and the self-similar
radiation collapse
described by the linear mass Vaidya space-time back ground,
showing that this radiation
collapse picture can be taken as a limiting case of  Tolman-Bondi
space-times when viewed in an appropriate sense.

It was shown in Ref.8 that if the positivity of energy was respected in the
near regions of the singularity, (i.e. $\epsilon + \it{P}>0$ in the
neighborhood of the singularity ) then an infinite many integral
curves terminate at the singularity which was naked. We show here
that a similar conclusion holds in the Tolman-Bondi case as well.

Let the energy density $\epsilon$ be positive in the collapsing region
near the central singularity at $r=0$, i.e.
$$ \epsilon = {\eta \Lambda \over R^2 H}>0\eqn\qq$$
This implies that $\Lambda_0>0$ and then the definition of $\eta$ implies
that $\alpha=\eta(0)$.
Let one simple positive root $X=X_0$ exist for the equation $V(X)=0$.
Note that in $(X,u)$ plane equation (30) has a singular point
at $X=X_0, u=0$. Therefore in order to analyze
the behavior of the integral curves
in $(X,u)$ plane near this singular point we
Integrate equation  (30) near the singularity to get
$$X -X_0= D u^{(h_0-1)}\eqn\qq$$
Hence in case $h_0<1$ integral curves move away from the singular point
$X=X_0,u=0$ in $(u,X) $ plane.
However, in the $(R,u)$ plane the above equation transforms to
$$R-X_0u=Du^{h_0}\eqn\qq$$
Therefore, if $h_0\le 0$ integral curves approaching
$R-X_0u$ in $(R,u)$ plane would move further and further from the
point $R=0,u=0$ and would not terminate there.
On the other hand, if $h_0>0$ the integral curves in $(R,u)$ plane
move into the point $R=0,u=0$ with either $R=X_0u$ or with the R-axis
as their ultimate tangent. In fact the equation of these integral
curves terminating at the singularity is given by
$${R\over u^{h_0}} -X_0u^{1-h_0}= D +\int S(Y,u)u^{(-h_0+1)}du\eqn\qq$$
where we have put
$$Y=R/u^{h_0}=Xu^{1-h_0}$$
and note that in the limit
$$\lim_{u\to 0} S(X,u)u^{1-h_0}= S(Y,u)u^{1-h_0}\to \hbox {const. }\times Y$$
Thus we see that
infinite many integral curves (each characterized by a different value
of the constant $D$) would terminate at the singularity
provided $h_0>0$.
Hence we deduce that future directed null geodesics would terminate
at the singularity in the past, as long as
$$\infty > h_o=h(X_o)> 0\eqn\qq$$
If positivity of energy in the near regions
of singularity is respected as stated in equation (42)
i.e. $\Lambda_0>0$, then  using equations (32) and (26) and the fact
that if $f(0)\neq 0$ then $\beta(0)=0$
we get for the value of $h_0$ when $\Lambda_0\neq 0$,
$$h_0={\Lambda_0H_0\over 2\alpha X_0^2(f_0+1)}\eqn\qq$$
Hence, we conclude that $h_0>0$ as long as the positivity of energy holds
in the near regions of the singularity.
Therefore families of geodesics would always
terminate at the singularity when it is naked and provided the positivity of
energy holds.

It is illustrative at this point to note the examples given in the earlier
section in the context of families meeting the singularity. Note that
for the first example given by equation (35),
in the case $n=2$ for example, $\Lambda_0=0$
and $V(X)=0$ has only one root given by $X=1$ and $h_0=1/2$.  Therefore
no families of
integral curves terminate at the singularity with the tangent $X=1$ . On
the other hand, for $n=1$, $\Lambda_0 \neq 0$ the space-time is self-similar
and
the families or infinitely
many non-spacelike curves terminate
at the singularity. The same is the case with the second example in
which $\Lambda_0 \neq 0$ where families would terminate at the singularity
when it is naked.

\vfil\eject

\centerline {C. Global Visibility}

It is  seen that the existence of a real positive root $V(X)=0$
establishes  that the singularity would be atleast locally naked. Such
a locally naked singularity could be globally naked as well. To examine
this issue
note that the apparent horizon lies at $R(t,r)=F(r)$, and therefore if a
geodesic gets inside the apparent horizon it becomes ingoing
(i.e. $R<F$ along geodesics and $dR/dr$ is negative)
Eventually this trajectory falls back to the singularity.
Therefore, if a light ray is to reach future infinity in order for the
singularity to be globally naked, it must cross
$r=r_c$, which is the boundary of the dust cloud before the apparent horizon.
Hence all escaping non-spacelike geodesics that reach the boundary $r=r_c$ with
$R(r_c)>F(r_c)$ would reach the future infinity. Since geodesics
emerge from the singularity with the tangent value $X_0$ and the apparent
horizon has the tangent at the singularity $\Lambda_0$, it follows
from equation (26) that $X_0>\Lambda_0$.
As a result,
because of the generality of the function $F(r)$ one can always choose
suitably $r_c$ and $F(r_c)=2M$ ($M$ being the Schwarzschild mass of the cloud)
such that geodesics reach the boundary of the cloud $r=r_c$ with
$R(r_c)>F(r_c)$ making the singularity globally naked. However, given
a boundary $r=r_c$ and $F(r_c)=2M$, which and whether any singular geodesics
would reach future infinity
depends on the global properties of the functions $F(r)$ and $f(r)$.

At this point we first discuss an explicit class of Tolman-Bondi models
where we show the singularity to be  globally naked,
before discussing the general scenario for global nakedness.
Due to the complicated nature of the equations,
exact solutions to geodesics are virtually non-existent in these models even
in cases of simple forms of functions $F(r)$ and $f(r)$,
except in the cases of Friedmann models corresponding to complete homogeneity.
We consider the first example given in the earlier section by equation (35)
for $n=1$.
This situation  represents a self-similar marginally bound collapse
($f=0$)
and illustrates the  formalism discussed here giving a comparison
with the results already obtained. Earlier, this
example has been analyzed using a special null trajectory which is
the Cauchy horizon [13] which is given by $X=const.$
We show below however, that actually one can integrate the geodesic
equations completely for this self-similar case to obtain radial null
families. As it was pointed
out  earlier, in this case if the condition
(38) is satisfied than $V(X)=0$ has two real positive and two complex
roots. Let $x_1$, $x_2$($x_1>x_2$) be two such positive roots of this equation.
Equation of geodesics, in the form $r=r(x), X=x^2$ is given by
$$r=r(X)\equiv r(x)=D{(x-x_2)^{n_2}\over (x-x_1)^{n_1}}f_1(x)\eqn\qq$$
where
$$f_1(x)=\exp(-\int{Ax+B\over x^2+D_1x+D_2}dx)\eqn\qq$$
Here $n_1,n_2,A,B,D_1,D_2$ are constants given by
$$x^4+{\sqrt{\Lambda_0}\over 2}x^3-x+\sqrt{\Lambda_0}
=(x-x_1)(x-x_2)(x^2+D_1x+D_2)\eqn\qq$$
$${3x^3 \over
x^4+{\sqrt{\Lambda_0}\over 2}x^3-x+\sqrt{\Lambda_0}}
={n_1\over x-x_1}-{n_2\over x-x_2}+{Ax+B\over x^2+D_1x+D_2}\eqn\qq$$
and $D$ is the constant which labels the different geodesics. The constants
$n_1,n_2$ are positive. In fact for the case
$\Lambda_0={7\over 17}$ they are given by
$$x_1=.658303,\quad x_2=.5,\quad n_1=2.09356,\quad n_2=1.08511,\quad
D_1=1.36419$$
$$D_2=1.2509,\quad
A=-1.99154,\quad B=-1.26354\eqn\qq$$
It is clear from equation (48)
that geodesics reach $r=0$ at $x=x_2$ and $r=\infty$ at $x=x_1$, making the
singularity globally naked. Note that $\eta(r)\Lambda(r)=F_0<x_2$
and therefore all the trajectories that are emitted in the region
$x_1>x>x_2$ reach the future infinity. In fact $x=x_1$ and $x=x_2$
are also geodesics which cross the boundary of the cloud and escape to
future infinity.

We now discuss  the conditions  which
ensure the global nakedness of the singularity in general.
At this point we assume  that the functions  $\eta$ and
$\beta$ are at least $C^0$ in the interval $r_c\ge r>0$. Since the later
two functions involve the first derivatives of $f$ and $F$ in the form
$f'/f$ and $F'/F$, this  requirement implies that $f$ and
$F$ have atleast first continuous derivatives existing.
As discussed in section 2,
the $C^2$-differentiability of the metric in  the
concerned interval will ensure the above
requirement.

Consider now the situation that $V(X)=0$  has only one simple root
$X=X_0$ and that a family of curves  terminates at the singularity (i.e.$h_0
>1$) with this value of tangent. Let $\eta(r)\Lambda (r)<\alpha X_0$
for $r_c\ge r>0$.
In such a situation the singularity would be globally naked.
To see this consider now the equation of geodesics given by equation (34)
where the constant $D$ labels different geodesics  terminating
at the singularity and is determined by the boundary conditions. For a
singular geodesic that
reaches the boundary of the dust cloud $u=u_c=r^{\alpha}=r_c^{\alpha}$ with
$X=(R_c/r^{\alpha}_c)=X_c$ we have
$$ X_c-X_0= Du_c^{h_0-1} +u_c^{h_0-1}\int_{u_c} Su^{(-h_0+1)}du\eqn\qq$$
and hence the equation of such a geodesic can be written as
$$ X-X_0= (X_c-X_0)({u\over u_c})^{h_0-1}
+u^{h_0-1}\int_{u_c}^{u} Su^{(-h_0+1)}du\eqn\qq$$
The event horizon is represented by the geodesic for which $X_c=\Lambda(r_c)$.
Since it is outgoing
$dR/d(r^{\alpha})$ is positive at $r=0$ and ejected into the region
$R>F$ where $dR/dr$ is positive. Therefore all the geodesics that
reach the line
$r=r_c$ (the line
at which the the metric (1) is matched with the Schwarzschild exterior)
with $X_c>\Lambda (r_c)$ would escape to infinity, while others would become
ingoing. It follows that the  geodesics that reach future infinity
with their past end point at the singularity are given by the equation
(54) with $X_c>\Lambda_c$.
Hence, in case when a family of geodesics  terminates at the singularity
with tangent $X=X_0$ and $\eta(r)\Lambda (r)<\alpha X_0$,
for $r_c\ge r>0$, the singularity
would be globally naked as
there would always be some geodesics that would escape to
infinity.

Consider the case now when the equation $V(X)=0$ has two positive roots
$X_1$ and $X_2$ ($X_1>X_2$). In such a case, as shown earlier, families
of curves would emerge from the singularity with the tangent either
$X_1$ or $X_2$. Let $\eta(r)\Lambda (r) <\alpha X_2$
for $r_c\ge r >0$, then it ensures that
some geodesics would cross the boundary of the cloud with $X_c>\Lambda(r_c)$
making the singularity globally naked. The same
holds even in case when more than two positive roots exist.
Thus if the family of geodesics do terminate at the singularity with
tangent $X_0$, then the condition $\eta(r)\Lambda (r)
<\alpha X_0$ for $r_c\ge r >0$ implies
the global nakedness of the singularity.

\vfil\eject

\beginsection 4. CURVATURE STRENGTH

Consider the case when naked singularities occurred in the
gravitational collapse of matter
with a reasonable equation of state and in a space-time where desirable
conditions such as the energy conditions etc. are satisfied. Even such a
situation may not be considered as a problem from the point of view of cosmic
censorship if the naked singularities forming were gravitationally weak
in a suitable sense. In fact it was shown [7] that the naked singularities
formi
   ng
in the classes of Tolman-Bondi models considered by Eardley and Smarr
and Christodoulou are gravitationally weak. This is a useful result,
because if true in general, it would have important implications for
the cosmic censorship hypothesis. Thus it was  conjectured that nature
avoids naked singularities where non-spacelike trajectories end in a strong
curvature singularity [7,22].

The gravitational strength and physical seriousness of a space-time
singularity have been discussed in detail and characterized precisely
in the literature.
In particular, Clarke and Krolak [23] have provided a sufficient condition
for a singularity to be strong in the sense of Tipler [24], which is that
atleast along one null geodesic with the affine parameter $k$, with $k=0$
at the singularity, the following should be satisfied
in the limit of approach to the singularity,

$$\lim_{k\rightarrow 0}k^2R_{ab}K^aK^b> 0\eqn\qq$$
This provides a sufficient condition for all the 2-forms $\mu(k)$
defined along the singular null geodesic to vanish as singularity is approached
and implies a very powerful curvature growth
establishing a strong curvature singularity.
For the timelike geodesics this will imply that all the volume forms defined
by the Jacobi fields along these trajectories  must vanish in the limit of
approach to the singularity or they must vanish infinitely many times in this
limit.

The criteria on the strength of a singularity are of course subject to further
refinement. However, the important physical consequences
of the existence of a singularity are
related to its strength. The point is if the singularity
is gravitationally weak, it may be possible to extend the space-time through
the same classically. On the other hand, when there is a strong curvature
singularity forming in the above sense, the gravitational tidal forces
associated with this singularity are so strong that any object trying to
cross it gets destroyed. Thus, as argued by Ori [25], the
extension of space-time becomes meaningless for such a strong singularity
which destroys to zero size all the objects terminating at the singularity.
{}From this point of view, the strength of singularity may be considered
crucial to the issue of classically extending the space-time and thus avoiding
the singularity; because for a strong curvature singularity
defined in the above sense,
no continuous extension of the space-time may be possible.

For the general class of Tolman-Bondi models under consideration,
using (2) we get
$$\Psi= R_{ab}K^aK^b={F'(K^t)^2\over R^2R'}={F'(K^t)^2\over R^2R'}
\eqn\qq$$
where $K^a$ is tangent to null geodesics.
For radial null geodesics,using L'hospital rule
and equations (4) to (14) and (19) to (22) and the fact that at the
singularity $r\to 0,X\to X_0$ we get
$$\lim_{k\rightarrow 0}k^2\Psi=\eta_0\lim_{k\to 0}
\left({k\sqrt{F}{\cal P}\over R^2\sqrt{rR'}}\right)^2=
{4\eta_0\Lambda_0\over H_0X_0^2
(2\sqrt{1+f_0}
(3\alpha -\eta_0)-N_0)^2}
\eqn\qq$$
Hence it is seen from the definition of $\Lambda$ in (12) that
$$\lim_{k\to 0}k^2\Psi =0\quad \hbox{for }\quad \alpha <\eta_0\eqn
\qq$$
$$\lim_{k\rightarrow 0}k^2\Psi \neq 0\quad \hbox{for }\quad \alpha \ge
\eta_0\eqn
\qq$$
However, from our earlier conclusions naked singularity occurs only when
$\alpha \le \eta_0$, therefore the strong curvature condition is satisfied
along singular geodesics only for the classes where $\alpha=\eta_0$.
As noted earlier, for the special class considered by Newman and
Christodoulou, $\alpha=7/3$ and $\eta=3$ and hence the naked singularity
turns out to be gravitationally weak as concluded earlier. On the other
hand, it is clear form the above that for a wide variety of Tolman-Bondi
solutions satisfying the condition $\alpha=\eta_0$, the singularity
will be a strong curvature singularity in the above sense.
In  general it is also possible that non-radial null
or timelike curves could  terminate
at the naked singularity. Then, a similar calculation along
non-spacelike geodesics
in general gives
$$\lim_{k\rightarrow 0}k^2\Psi \propto \left(r^{\eta_0-\alpha}
\right)_{r=0}\eqn\qq$$
Hence as discussed above one concludes that condition for strong curvature
is satisfied along non-spacelike geodesics as well if $\alpha =\eta_0$
and if such families meet the naked singularity in the past.

The Kretschmann scalar $R_{abcd}R^{abcd}$ along the geodesics
goes in the Tolman-Bondi space-times as
$${\cal K}\propto  r^{2(\eta_0-3\alpha)}\eqn\qq$$
Hence the singularity is a scalar polynomial singularity as long as $\alpha
>{\eta_0\over 3}$.

The self-similar Tolman-Bondi models are defined by the conditions
$f(r)=const.$ and $\eta(r)=1=\eta(0)=\alpha$. It follows from the above that
the naked singularity forming in this class will be a strong curvature
singularity along all the families of radial null geodesics. As shown in
Ref.8, other families of non-spacelike geodesics also do terminate
at the naked singularity along which as well the strong curvature
condition is satisfied.

\vfil\eject

\beginsection 5. CONCLUDING REMARKS

We have analyzed here the Tolman-Bondi models for the
existence and structure of the naked singularities. As  stated
earlier, these are dust models  assuming the pressure
$p=0$,  and
also the exact spherical symmetry of the space-time.
Would the introduction of pressure change the qualitative nature of
the conclusions obtained here ? This does not seem to be the case
atleast for the  self-similar gravitational collapse of a perfect fluid
incorporating pressure as indicatd by the analysis of [14] and [8].
It is possible, on the other hand, that in the final stages of collapse, the
dust equation of state could be  relevant (see e.g. Penrose [26],
Hagerdorn [27]) and at higher and higher densities the matter may behave
more and more like dust.
Again, there is some case for the argument that eventually in the final
stages of collapse, the matter distribution should become almost
spherically symmetric (see e.g. Nakamura and Sato [28] ). Hence, it
is clearly useful to examine the inhomogeneous dust collapse as modelled
by the Tolman-Bondi space-times.
Further, a situation analogous to the singularity theorems
might develop here where the conclusions derived under the assumption of
spherical symmetry are preserved when small perturbations are taken into
account. Thus, spherical symmetry may be a good model to represent a certain
class of gravitational collapse.

Also, we have not addressed
the issue of the stability of  naked singularity. If these are not stable
(in a sense to be defined suitably) such naked singularities need not be
considered as counter-examples to the cosmic censorship hypothesis.
As far as the issue of stability is concerned,
one needs to develop a precise criterion for stability in general relativity.
In this connection it may be noted however, that for self-similar Tolman-Bondi
models the Cauchy horizon is stable atleast against the blue shift mode of
instability [21].

Subject to these reservations, it is seen here that
Tolman-Bondi space-times  admit
naked singularities under fairly general conditions,
from which a non-zero measure set of non-spacelike trajectories
emanate in the future direction.
Certain examples of particular classes where
non-radial non-spacelike geodesics  terminate at the naked singularity
in the past are also explicitly worked out.
An interesting point is that in the case of the
strong curvature condition being satisfied along radial null trajectories,
the same conclusion also holds along all other non-spacelike geodesics.
For various other classes of naked singularity space-times, even though
the strong curvature condition  may not be satisfied
along radial curves, they
could still be regarded as strong curvature singularity in the sense that
the Kretschmann scalar diverges.

Another feature one would like to note here is  that
while   strong curvature naked
singularities have been found to occur in self-similar gravitational
collapse as indicated earlier,
the present consideration gives a  wide class of inhomogeneous collapse
models which need not be
self-similar in general.
A wide class of space-times has been pointed out,
namely the ones for which $\alpha=\eta_0$, which gives
a set of solutions of the field equations which
admit a strong curvature naked singularity. The suggestion that seems to be
coming is that
the phenomena of naked singularity is probably not related to
the space-times with any particular geometric properties such as the
self-similarity of the models.
It may be that the existence of naked singularity
is not just a geometric phenomena and the answer to cosmic censorship
conjecture could lie in the dynamics of the Einstein equations. Of course,
if one rules out the matter fields such as the dust and perfect fluid etc.
from consideration because they may create singularity even without gravity,
then such naked singularities are ruled out (see however, [29] where the
occurrence of naked singularity is pointed out for a wide range of matter
satisfying the weak energy condition in self-similar gravitational
collapse).

To summarize, the conclusions on the final fate of gravitational
collapse are rather different in the generally inhomogeneous Tolman-Bondi
models as compared to the  Oppenheimer-Snyder case of
a completely homogeneous dust collapse, which forms a set of zero measure in
the general Tolman-Bondi class considered  here. In fact, the similarity
in conclusions concerning the nature and structure of the naked singularity
for the radiation collapse [20], the general self-similar
collapse [8] of perfect fluid, and the results here appear suggestive of
a certain general property of Einstein equations. It would be worth while to
isolate and study this feature as that might help towards a definite
mathematical formulation of the cosmic censorship by pointing out the
precise feature  one wants to rule out. Such a study would be of
independent interest any way because not much is understood on the global
properties of the Einstein equations except the results on the existence
of space-time singularities as predicted by the singularity theorems.

While the analysis we have presented here should be useful
towards arriving at any rigorous formulation of cosmic censorship in a
provable form as pointed out above,
we would like to argue here
that a physical formulation of the cosmic censorship may be evolved
which avoids features such as above.
For example, an interesting feature that  emerges from  the
presently available examples is the role of energy conditions in
determining the escape of families of non-spacelike trajectories from the
naked singularity,  which is an important criteria
for the physical significance
of the same. In all the presently available collapse scenarios,
it is the weak energy condition together with the positivity of
of energy, which leads to the existence of families
of non-spacelike geodesics  terminating at the naked singularity in the past.
Could one then argue that some how the energy conditions must be
violated in the very final stages of gravitational collapse so as to
avoid the formation of naked singularity ? In fact, in the case of
self-similar collapse [8], it can be shown that the violation of the energy
condition near the singularity no longer allow the families of
non-spacelike geodesics to come out but only an isolated trajectory can
emerge. Hence, for all practical purposes, the singularity is no longer
naked preserving the effective censorship. Again, as emphasized by
Israel [30], many of the naked singularities arising in the spherically
symmetric collapse are massless (with the mass  being defined
in a suitable manner, see also Lake [18]); and as a consequence these
may not violate the basic physical spirit of the cosmic censorship.
Such possibilities need a serious investigation.

\vfil\eject

\centerline {$\underline {\cal REFERENCES}$}

\item{\bf 1.} R.C.Tolman,  Proc. Natl. Acad Sci USA 20 410 (1934).
\item{\bf 2.} H. Bondi,  Mon. Not. Astron. Soc. 107 343 (1947).
\item{\bf 3.} J. Oppenheimer  and H. Snyder,  Phys.Rev. 56, 455(1939).
\item{\bf 4.} R. Penrose, Riv. Nuovo Cimento Soc. Ital. Fis. 1, 252
(1969); for further discussion see e.g. R.Penrose in "General Relativity,
an Einstein centenary survey" ed. S.W.Hawking and W. Israel (1979),
Cambridge Univ. Press, and also
W. Israel, Found. Phys. 14, 1049(1984); Can.J.Phys. 64, 120(1986).
\item{\bf 5.} D. M. Eardley and L. Smarr, Phys. Rev. D19,
2239 (1979); see also D. M. Eardley in "Gravitation in Astrophysics", Plenum
Publishing Corporation Edited by B. Carter and J. B. Hartle 223 (1987).
\item{\bf 6.} D. Christodoulou, Commun Math. Phys. 93 , 171 (1984).
\item{\bf 7.} R.P.A.C. Newman, Class. Quantum Grav. 3, 527 (1986).
\item{\bf 8.} P.S.Joshi and I.H.Dwivedi, Commun. Math. Phys. 146,333(1992).
\item{\bf 9.} This supports the earlier conclusions derived in the
study of radiation collapse models, especially when the collapse is
not self-similar; see e.g. Y.Kuroda, Prog.Theor.Phys. 72, 63(1984),
K.Rajagopal and K.Lake, Phys.Rev. D35, 1531(1987),
P.S.Joshi and I.H.Dwivedi, J.Math.Phys. 32, 2167(1991),
K.Lake, Phys.Rev. D43, 1416(1991),
P.S.Joshi and I.H.Dwivedi, Gen.Relat.Grav. 24, 129(1992), Phys.Rev. D45,
2147(1992).
\item{\bf 10.} See e.g. D.M.Eardley(1987) in [5].
The Tolman-Bondi models are fully characterized in terms
of the behavior of
the functions $f$ and $F$ and hence the conditions for the absence of shell
crossing singularities for $r>0$ can be given in terms of the behavior of
these functions in the range $r_c\ge r>0$ ($r_c$ being the boundary
of the dust cloud). For further details and conditions see e.g.
J. Hellaby and K.Lake, Astrophys. J. 290, 381(1985),
R.P.A.C. Newman in Ref[7].
\item{\bf 11.} H. Muller, P. Yodzis and H. Seifert, Commun. Math. Phys,
37 29-40 (1974); see also Ref. 10 and 12.
\item{\bf 12} A.Papapetrou and A.Hamoui , Ann. Inst. H Poincare
VI 343,(1967). We note however, that such an extension need not be unique
or even be dust.
\item{\bf 13} B.Waugh and K.Lake, Phys. Rev. D38, 4, 1315 (1988).
\item{\bf 14} A.Ori and T.Piran, Phys. Rev. D42, 4, 1068 (1990).
\item{\bf 15} G.Grillo, Class. Quantum Grav. 8, 739 (1991).
\item{\bf 16} I.H.Dwivedi and P.S.Joshi, Class. Quantum Grav.
9, L69 (1992).
\item{\bf 17} If one requires the metric to be
$C^2$ from the point of view of physical reasonableness,
then the differentiability conditions
imposed here on the functions
$f$ and $F$ should be considered reasonable.
Even when one requires the weaker condition that the metric is
finite and continuous at the center, the functions $f$ and $F$ must
atleast be $C^1$ at $r=0$ (though the converse is not true). Further, from
the physical resonableness one would require $F(0)=0$ (otherwise there
will be a massive singularity present already at $r=0$), which implies
$\eta(0)>0$. The condition $F$ being $C^1$ corresponds to the energy
density $\epsilon$  not diverging at the center $r=0$ at all times.
\item{\bf 18} See e.g. the discussion on the nature of the naked singularity
in such a case by K.Lake, Phys.Rev.Lett. 68, 3129(1992).
One could reduce the right hand side of the
equation (23) for the marginally
bound case $f=0$ and for radial null geodesics ($\ell=0, B=0$) to the
form $P/Q$, where both $P$ and $Q$ vanish at $r=0,R=0$.
However, even in this simplified case
$P$ and $Q$ have no linear terms  and
so the analysis on the nature of the singularity by the usual techniques
(see e.g. T.Davies and E.James, "Non-linear Differential Equations" Addison
Wesley NY(1966) or F. Verhulst, "Non-linear Differential Equations and
Dynemical Systems", Springer-Verlag, Berlin,1990 )  do not apply.
The only way then to understand the nature of singularity
is a direct analysis of the behavior of the characteristic
curves in the neighborhood of the singularity, which we have done here.
\item{\bf 19} The tangent to the apperent horizon $R=F$ in $(R,u)$ plane
is $(\eta \Lambda/\alpha)$ and at the singularity it becomes $(\eta(0)\Lambda(
0)/\alpha)$ which is effectively given by $\Lambda(0)$. This is because
when $\alpha <\eta(0)$ then $\Lambda(0)=0$ and when $\alpha = \eta(0)$ the
value of tangent becomes $\Lambda(0)$.
\item{\bf 20.} I.H. Dwivedi and P.S. Joshi, Class. Quantum Grav.
6, 1599 (1989); 8, 1339 (1991). The nature and structure of the families of
non-spacelike geodesics coming out of the naked singularity, forming in
the self-similar radiation collapse (with a linear mass function), and the
directional nature of the curvature growth along the same is examined here.
\item{\bf 21.} J.P.S. Lemos, Phys.Rev.Lett. 68, 1447(1992).
\item{\bf 22.} F. J. Tipler, C. J. S. Clarke and G. F. R. Ellis 1980 'General
Relativity and Gravitation' Vol 2, ed A Held (New York : Plenum) p 97.
\item{\bf 23.} C.J.S.Clarke and A.Krolak, J. Geo. Phys. 2, 127 (1986).
\item{\bf 24.} F.J.Tipler, Phys. Lett. 64A, 8(1977).
\item{\bf 25.} A.Ori, Phys.Rev.Lett. 67, 789(1992).
\item{\bf 26.} R.Penrose, in "Gravitational Radiation and Gravitational
Collapse" (IAU Symposium No.64) ed. C.DeWitt-Morettee, Reidel, Dordrecht(1974).
\item{\bf 27.} R.Hagerdorn, Nuovo Cimento 56A, 1027(1968).
\item{\bf 28.} T.Nakamura and H.Sato, Prog.Theor.Phys. 67, 346(1982).
\item{\bf 29.} P.S.Joshi and I.H.Dwivedi (1993), to appear in
Lett.Math.Phys.
\item{\bf 30.} W.Israel (1992), private communication.

\end